\input harvmac
\noblackbox
\lref\kmm{
D.~Kutasov, M.~Marino and G.~W.~Moore,
``Remarks on tachyon condensation in superstring field theory,''
hep-th/0010108.}
\lref\rab{See e.g. 
S.~A.~Abel, J.~L.~Barbon, I.~I.~Kogan and E.~Rabinovici,
``String thermodynamics in D-brane backgrounds,''
JHEP {\bf 9904}, 015 (1999)
hep-th/9902058.}
\lref\zam{
V.~Fateev, A.~B.~Zamolodchikov and A.~B.~Zamolodchikov,
``Boundary Liouville field theory. I: Boundary state and boundary  two-point function,''
hep-th/0001012.}
\lref\mgas{M.~Gutperle and A.~Strominger,
``Spacelike branes,''
JHEP {\bf 0204}, 018 (2002) hep-th/0202210.}

\lref\hull{C.~M.~Hull,
``Timelike T-duality, de Sitter space, large N gauge theories and  topological field theory,''
JHEP {\bf 9807}, 021 (1998) hep-th/9806146.}

\lref\SenVV{A.~Sen,
``Time evolution in open string theory,''
hep-th/0207105.}

\lref\SenAN{A.~Sen,
``Field theory of tachyon matter,''
hep-th/0204143.}

\lref\SenIN{ A.~Sen,
``Tachyon matter,''
hep-th/0203265.}

\lref\SenNU{A.~Sen,``Rolling tachyon,''
JHEP {\bf 0204}, 048 (2002)
hep-th/0203211.}

\lref\tescb{
B.~Ponsot and J.~Teschner,
``Boundary Liouville field theory: Boundary three point function,''
Nucl.\ Phys.\ B {\bf 622}, 309 (2002) hep-th/0110244.}

\lref\tesca{
J.~Teschner, ``Remarks on Liouville theory with boundary,''
hep-th/0009138.}
\lref\Braa{
E.~Braaten, T.~Curtright, G.~Ghandour and C.~B.~Thorn,
``Nonperturbative Weak Coupling Analysis Of The Liouville Quantum Field Theory,''
Phys.\ Rev.\ Lett.\  {\bf 51}, 19 (1983);
Annals Phys.\  {\bf 153}, 147 (1984).}

\lref\inp{M. Gutperle, A. Recknagel, V. Schomerus and A. Strominger, in 
progress.}
\lref\gpas{M. Gutperle and A. Strominger, unpublished.}
\lref\SenMG{
A.~Sen,
``Non-BPS states and branes in string theory,''
hep-th/9904207.}
\lref\jp{
J.~Polchinski,
``Remarks On The Liouville Field Theory,''
UTTG-19-90
{\it Presented at Strings '90 Conf., College Station, TX, Mar 12-17, 1990}.
}
\lref\Okuda{
T.~Okuda and S.~Sugimoto,
``Coupling of rolling tachyon to closed strings,''
hep-th/0208196.}
\lref\Deger{
N.~S.~Deger and A.~Kaya,
``Intersecting S-brane solutions of D = 11 supergravity,''
JHEP {\bf 0207}, 038 (2002), hep-th/0206057.}
\lref\Roy{
S.~Roy,
``On supergravity solutions of space-like Dp-branes,''
JHEP {\bf 0208}, 025 (2002), hep-th/0205198.}
\lref\Myers{
M.~Kruczenski, R.~C.~Myers and A.~W.~Peet,
``Supergravity S-branes,''
JHEP {\bf 0205}, 039 (2002), hep-th/0204144.}
\lref\Chen{
C.~M.~Chen, D.~V.~Gal'tsov and M.~Gutperle,
``S-brane solutions in supergravity theories,''
Phys.\ Rev.\ D {\bf 66}, 024043 (2002), hep-th/0204071.}
\lref\Lu{
H.~Lu, S.~Mukherji and C.~N.~Pope,
``From p-branes to cosmology,''
Int.\ J.\ Mod.\ Phys.\ A {\bf 14}, 4121 (1999)
hep-th/9612224.}
\lref\Lub{
H.~Lu, S.~Mukherji, C.~N.~Pope and K.~W.~Xu,
``Cosmological solutions in string theories,''
Phys.\ Rev.\ D {\bf 55}, 7926 (1997), hep-th/9610107.}

\def\p{\partial}
\def\o{{\omega }}
\def\apm{{\alpha^\prime}}


\Title{\vbox{\baselineskip12pt\hbox{hep-th/0209090}\hbox{}
\hbox{}}}{Open String Creation by S-branes }

\centerline{Andrew Strominger }
\bigskip\centerline{Department of Physics}
\centerline{Harvard University}
\centerline{Cambridge, MA 02138}

\vskip .3in \centerline{\bf Abstract}
 An sp-brane can be viewed as the creation and decay of an unstable 
D(p+1)-brane. It is argued that the decaying half of an sp-brane can 
be described by a variant of boundary Liouville theory.  The pair creation
of open strings by a decaying s-brane is studied in the minisuperspace
approximation to the Liouville theory. In this approximation a
Hagedorn-like divergence is found in the pair creation rate, suggesting
the  s-brane energy is rapidly transferred into closed string radiation.

\bigskip 
\smallskip
\it{Talk 
presented at the String 2002 Conference, August 12-15, Hangzhou, China}
\Date{}
\listtoc
\writetoc

\newsec{Introduction}

 An s-brane (spacelike-brane) is a topological defect all of whose 
longitudinal dimensions are spacelike and therefore exists only for a
moment in time. The topological ``s-charge'' is typically defined 
by an integral over a sphere which has both timelike and spacelike
directions.  In \mgas\ it was argued that string theory contains 
s-branes. (See also \refs{\hull \Lu \Lub \Chen \Myers \Deger \Roy 
-\Okuda}.) The argument invoked the picture \SenMG\ of stable Dp-branes 
as domain walls spatially interpolating between two minima of 
the tachyon potential of unstable D(p+1)-branes. Sp-branes are 
related topological configurations in which the interpolation occurs in 
the time, rather than space, direction. Energy conservation requires both 
incoming and outgoing matter/radiation  of some kind. 
An sp-brane can be described as a
process in which incoming  matter/radiation excites the tachyon field up to 
the top of the potential, from where it subsequently 
decays into the next minimum. 

S-branes provide an interesting and relatively simple arena in which 
to study time-dependent processes in 
string theory. In the large N limit, with 
N the number of S-branes, they may be holographically dual to interesting 
closed string cosmologies. 

Recently Sen \refs{\SenNU \SenIN \SenAN -\SenVV} has argued that a 
worldsheet boundary interaction 
$\sinh X^0$, where $X^0$ is the timelike coordinate, gives an 
exact conformally invariant boundary 
sinh-Gordon type field theory and constructed
the corresponding boundary state. This amounts to a CFT construction 
of the s-brane in a formal $g_s = 0$ limit, in which quantum effects and
closed strings are suppressed. It may be checked from the RR boundary state 
guessed in \SenAN\ that the s-brane indeed carries 
the required RR s-charge \gpas. 

This talk we will discuss the related $e^{X^0}$ interaction, which is
a kind of boundary Liouville theory for negative norm bosons. 
(Ordinary boundary Liouville theories have been studied in \refs{\zam 
\tesca -\tescb}.) This is formally a limit of the $\sinh (X^0-a)$ type theory in which
the location $a$ of the s-brane is taken to past infinity while scaling the
interaction strength. It describes only the future, decaying, half of the solution, 
or a half-s-brane. 

All of these solutions have the property, as shown in \SenNU,  that in 
the far future, after the decay is completed, the energy of the s-brane is
converted into a pressureless tachyon dust which remains confined to the 
unstable D(p+1)-brane worldvolume. This tachyon dust 
is somewhat bizarre because 
the tachyon has reverted to its minimum and there should be no open string
excitations. Presumably this picture is greatly modified for 
$g_s\neq 0$ \Okuda.  Away from $g_s=0$, one expects that the energy should leak off
the brane in the form of closed string radiation. 

This talk will report on work in progress analyzing the 
decay process in bosonic string 
theory for $g_s \neq 0$. As a first step, in order to get a picture
of  what to expect,  an adaptation of the 
'mini-superspace' approximation of \Braa\ (see also \jp) is employed. 
This
approximation 
has proven
generally useful for ordinary Liouville theory, but its validity has not
been demonstrated in the present context. 
As the problem considered here is comparable in 
complexity to those solved in  \refs{\zam 
\tesca -\tescb}, a more systematic analysis should be possible. We hope to
report on this in the near future \inp. 

In the minisuperspace approximation all open string
states acquire exponentially growing masses from the tachyon background. 
This results in open string pair
production.
We will find that for any nonzero $g_s$, the pair production
is typically divergent\foot{The divergence can apparently 
be suppressed for 
large numbers of noncompact transverse dimensions \rab, as discussed in the
last section.}  and the theory becomes strongly coupled in a time 
of order $(g_s)^0$. Roughly speaking what happens is that the decaying
brane tries to produce open strings at the Hagedorn temperature. 
Whether or not this behavior survives the minisuperspace approximation
remains to be seen \inp. 
This suggests that the $g_s \to 0$ limit is not smooth. A divergent density
of open strings (with exponentially growing masses) 
will couple strongly to closed strings, suggesting
that the s-brane energy is quickly released into closed strings.   
However string perturbation theory probably 
cannot be used to follow this process
all the way through because of the intermediate strong coupling region.

The paper is organized as follows. In section 2 we compute the particle
creation rate for a scalar field with an exponentially growing mass. 
In section 3 we show that in the minisuperspace approximation the effect of
a time-dependent background tachyon is to give time-dependent masses to the 
open string modes. In section 4 we give the description of the half-s-brane 
as a boundary Liouville theory. In section 5 we put this together and
estimate the total perturbative open string creation rate. We find that it 
typically diverges due
to the large number of states at high energy. Implications are discussed.

\newsec{Scalar Field with Time-Dependent Mass}
  Consider the Klein-Gordon equation for a complex scalar
  with a time-dependent mass
\eqn\ftyu{(-\p_t^2+\nabla^2)\phi=m^2(t)\phi,} 
\eqn\rrd{-\p_t^2\phi=(m^2(t)+\vec p^2)\phi.} For sufficiently slowly
varying $m$ the solutions behave as  \eqn\sare{\phi \sim e^{\pm i
\sqrt{m^2(t)+\vec p^2}t},}and hence oscillate more rapidly as $m$ is 
increased. The Klein-Gordon
current \eqn\kgc{j_a=i(\phi^*\p_a\phi-\phi\p_a\phi^*) } is conserved for any
$m(t)$.  However since the mass is time dependent energy will not
be conserved: raising the mass requires an input of energy.

\subsec{Exponential Mass} Consider the special case
\eqn\jkl{m^2(t)=e^{2 t}+m_0^2.} 
We shall see that this is the typical behavior of an open string mass 
in a half-s-brane background: the mass goes to infinity for 
$t \to \infty$ as the tachyon
decays.   
The solutions of \rrd\ with \jkl\ are 
\eqn\iio{\psi^{in}_{\vec p} ={ 2^{-i\o} \over \sqrt{2 \o}}\Gamma(1-i\o )
e^{i\vec p \cdot \vec x}    J_{-i\o}(e^t),~~~~\omega
\equiv \sqrt{m_0^2+\vec p^2}} and its complex conjugate. These 
are normalized so that 
\eqn\nrm{ j_0(\vec x,t)=1.}
In the far past
this solution approaches a positive frequency plane wave 
\eqn\jfh{t\to
-\infty,~~~~~~\psi_{\vec p}^{in}\to {1 \over \sqrt{2\o}}e^{-i\o t+i\vec p \cdot \vec x} .}
In the far future 
\eqn\zjfh{t\to
\infty,~~~~~~\psi_{\vec p}^{in}\to { 2^{-i\o} \Gamma(1-i\o )\over \sqrt{\pi  \o}}
e^{- t/2+i\vec p \cdot \vec x}\cosh({\pi \o \over 2}-ie^t+i {\pi \over 4}) .}
Note that the frequency increases exponentially, corresponding to the
higher energy, while the amplitude decreases. 

\subsec{Hyperbolic Mass}
Another relevant special case is $m^2(t)=\cosh 2t$:
\eqn\rft{\p_t^2\phi+(\cosh 2t +\vec p^2)\phi=0.}
This arises for open strings in a full s-brane, which are very massive 
in both the far past and the far future. 
The solution to this equation is a Mathieu function with an imaginary 
argument, which for large values
of the argument approaches a Bessel function.  

\subsec{Particle Creation}

According to \zjfh, for an exponentially growing mass,  
the incoming modes $\psi_{\vec p}^{in}$ contain both negative and positive
frequency parts in the far future. This indicates particle production. 
Normalized outgoing positive frequency modes are Hankel functions
\eqn\otm{\psi_{\vec p}^{out}= \sqrt{\pi \over 4 i}e^{-{\pi \o \over 2}
+i\vec p \cdot \vec x} H_{-i\o}^{(2)}(e^t)
\to {1 \over \sqrt{2}}e^{-{t \over 2}-ie^t+i\vec p \cdot \vec x},~~~~t \to
\infty.}
The energy of these modes in the far future goes as 
\eqn\enrg{{\cal E}(t)=\dot \psi_{\vec p}^{out*}\dot \psi_{\vec p}^{out}
+(\omega^2 +e^{2t})\psi_{\vec p}^{out*}\psi_{\vec p}^{out} \sim e^t\sim m(t).}
The energy grows with time because the equation of motion is time 
dependent. 
Using
\eqn\dfl{\eqalign{ H_{-i\o}^{(2)}(x)&=
{1 \over \sinh \pi\o}\bigl( e^{\pi\o}J_{-i\o}(x)-J_{i\o}(x) \bigr) ,\cr
J^*_{i\o}(x)&=J_{-i\o}(x),\cr
J_{-i\o}(x)&=\half(H^{(2)}_{-i\o}(x)+H^{(2)*}_{i\o}(x))  ,\cr
|\Gamma(1+i\o)|^2&={\pi \o \over \sinh \pi \o },}}
for real $x,~\o$, we find
\eqn\fda{\eqalign{\psi_{\vec p}^{out}&=\alpha_{\vec p} \psi^{in}_{\vec p}
+\beta_{\vec p}\psi^{in*}_{-\vec p},\cr
\psi_{\vec p}^{in}&=\alpha_{\vec p}^* \psi^{out}_{\vec p} -\beta_{\vec p}\psi^{out*}_{-\vec p},\cr
 \alpha_{\vec p}&=  {2^{i\o}  \over\sqrt {2 \pi i \o}} \Gamma (1+i\o)e^{\pi \o \over 2}, \cr
\beta_{\vec p} &=-{ 2^{-i\o}\over\sqrt {2 \pi i \o}}
\Gamma (1-i\o)e^{-{\pi\o \over 2} }.}}
Note that these Bogolubov coefficients 
obey $\alpha_{\vec p}\alpha^*_{\vec p}-\beta_{\vec p}\beta^*_{\vec p}=1$ as
required.
For large $\o$, $\beta_{\vec p}  $ approaches zero while $\alpha_{\vec p} $ becomes a 
pure phase. Expanding
\eqn\exp{\phi=\sum_{\vec p} \bigl(\psi^{in}_{\vec p} a^{in}_{\vec p} +\psi^{in*}_{\vec p
}a_{\vec p}^{in\dagger}\bigr)=\sum_{\vec p} \bigl(\psi^{out}_{\vec p} a^{out}_{\vec p} +\psi^{out*}_{\vec p
}a_{\vec p}^{out\dagger}\bigr),}
the in vacuum becomes
\eqn\ffd{|in>=\prod_{\vec p} (1-|\gamma_{\vec p}|^2)^{1/4} e^{-\half\sum \gamma_{\vec p} (a_{\vec p}^{out\dagger})^2}|out>,}
where
\eqn\wsx{\gamma_{\vec p}={\beta_{\vec p}^* \over  \alpha_{\vec p}}= -ie^{-\pi \o}.}
The in vacuum is annihilated by $a_{\vec p}^{in}=\alpha_{\vec p} a_{\vec p}^{out}+\beta^*_{\vec p} 
a_{-\vec p}^{out\dagger}.$

Relation \ffd\ expresses the fact that if there are no incoming particles
at $t\to -\infty$, there will necessarily be outgoing particles 
at $t \to \infty$. The expected total number of 
particles created in each mode $\vec p$ is finite
and largely occurs in a neighborhood of $t=0$. The creation 
is exponentially suppressed at frequencies 
$\omega >> {1 \over \pi}$.

\newsec{Minisuperspace Approximation}

In this section we consider the quantization of an open string 
in the half-s-brane using the minisuperspace approximation for the 
zero mode of $X^0$. This follows analogous treatments for the 
ordinary bulk Liouville theory in \refs{\Braa,\jp}.

The bosonic part of the worldsheet action for an open string on an
unstable bosonic D-brane is \eqn\lij{-{1 \over 4\pi\apm}\int d\tau d\sigma\sqrt{-\gamma}\gamma^{ab} \p_a X^\mu
 \p_b X_\mu -{1 \over 8\pi} \int d\tau\sqrt{-h} T(X),} where $T$ is the
background tachyon field, $0\le \sigma \le \pi$  and $h$ is 
the boundary metric induced from the
worldsheet metric $\gamma$. The second integral runs over all boundaries of
the worldsheet. For the superstring 
$T$ is replaced by $T^2$ \refs{\SenMG,\kmm}, ensuring that the potential is 
bounded from below for 
all background tachyon fields. 

In a minisuperspace type 
approximation we treat the zero mode as independent of 
the oscillators. We wish to consider 
the zero mode behavior for 
the case $T=T(X^0)$.  The pure 
zero mode piece of \lij\ for $\gamma=\eta$ is (for $X^\mu$ independent of 
$\sigma$)
\eqn\gxz{ S_0= {1 \over 4\apm}\int d\tau\bigl(-(\dot X^0)^2+(\dot
X^i)^2 -{\apm T(X^0)\over \pi}\bigr).} The corresponding zero mode Hamiltonian is
\eqn\ham{H_0=-\apm (P^0)^2+\apm (  P^i)^2
+{T(X^0)\over 4 \pi} .} Including oscillator contributions 
the constraint $L_0+\bar L_0=0$ 
becomes the Schroedinger
equation for the open string wave function \eqn\seql{\bigl( {\p^2
\over \p X^{02}} +{T(X^0)\over 4 \pi\apm}+{N-1 \over \apm}+ p^ip^i \bigr)\psi(X^0)=0.} Here
$p^i$ are the constants of motion corresponding to the zero mode momenta 
$P^i$ in \ham\ and  $N$ are  the ``oscillator'' contributions.

Making the identification $t \to X^0$ and 
\eqn\idt{m^2(X^0)+\vec p^2= {T(X^0)\over 4 \pi\apm}+{N-1 \over \apm}+ p^ip^i     }
\seql\ is in fact exactly the wave equation 
\rrd\ for a scalar field with time-dependent mass. This might have 
been anticipated. The worldsheet tachyon interaction is represented by 
the proper length of the end of the string worldsheet weighted by the 
background tachyon. This is equivalent to hanging masses at the ends of 
the string proportional to $T(X^0)$. Therefore a time dependent tachyon
gives a time-dependent mass to all open string states. In the limit 
of $T \to \infty$ one expects to recover the closed string vacuum. Indeed
in this limit all open string masses go to infinity.

\newsec{The Half-s-brane and Boundary Liouville Theory}
In order that \lij\ describe a consistent open string background 
$T$ should be a conformally invariant interaction.  We consider
the case 
\eqn\fdt{T=e^{X^0/\sqrt{\apm}}.}
This is a dimension one operator because the $X^0$ has the 
``wrong sign'' two-point function. 
\fdt\ defines a boundary Liouville theory, which has been studied 
(for a positive norm field) in  \refs{ \zam \tesca -\tescb}.
In the past $X^0 \to -\infty$,  the tachyon field is perched at the 
top of its potential. It falls off towards the future, and reaches the 
closed string vacuum in the far future, where all open strings become 
infinitely massive. It describes the decay of an unstable brane or, 
equivalently, the future half of an s-brane.

The open string spectrum is of course continuous because of the noncompact 
region. In the usual (positive norm) 
boundary Liouville theory there is a non-trivial 
density of states which is 
related in \tesca\ to the reflection coefficients off of the tachyon
wall, which in turn are related to the two point function on the disc.
In our case however, there is no reflection simply because the 
potential implied by \fdt\ has a negative sign relative to the $X^0$ kinetic term, and
both solutions are allowed in the future. Hence 
the density of states is the same as it is for free field theory. 

This has a simple interpretation. The open string spectrum on the half 
s-brane can be computed in the far past $X^0\to - \infty$ where the effect
of the tachyon is negligible. The $L_0$ eigenvalue does not depend on 
$X^0$ and so is the same in the far future. However the open string 
eigenfunctions are of course functions of $X^0$. In fact the effective open
string masses appearing in \idt\ are of the exponential form \jkl\
\eqn\expl{m^2(X^0)=m_0^2+{e^{X^0/\sqrt{\apm}}\over 4\pi\apm},}
for $X^0=2\sqrt\apm t$.
Hence the typical behavior of the open string eigenfunctions is 
expected to be of the form \iio\ with rapid oscillations in the far
future.

\newsec{Hagedorn Divergence}

In this section we discuss open string pair production on a half-s-brane.
The first step is to define the incoming state of the open string modes.
Since the background tachyon 
field vanishes in the far past, it is natural to take all the incoming 
open string 
modes in their usual ground state. This does not work for the tachyon for
the usual reason: there is no quantum ground state for a particle perched
on top of a hill. If we try to force it to sit exactly at the maximum,
it will have large fluctuations in its momentum. What this means is that 
in the quantum world the decay process will always begin some finite 
time after the initial conditions are set. In the following we presume 
that the creation of non-tachyonic open strings modes, whenever this decay 
process starts, is approximated by open string creation in 
the half-s-brane background.  

The (expectation value of the) energy dumped by the decaying brane 
into open string pair production will contain 
factors such as 
\eqn\rlp{\int N_{\omega}d{\cal E}_{\omega}.}
The sum is over all single open string modes with incoming energy labelled by 
$\omega$ and density of states $N_{\omega}$. $d{\cal E}_\o$ is the 
(expectation value of the) outgoing energy in these modes, which may
contain multiple quanta.  
Approximating the outgoing state 
by the right hand side of \ffd\ and using \enrg\ one finds for large $t$
\eqn\trr{d{\cal E}_\o={e^{2 \pi X^0  T_H} d^p{\vec p} \over (2 \pi)^p( e^{\o/T_H }-1)}, 
~~~~~T_H={1 \over 4 \pi \sqrt{\alpha'}}   ,}
where $p$ is the dimension of the s-brane.\foot{Frequencies here 
are defined relative to $X^0=2\sqrt\apm t$.}  At large $\omega$ \rab 
\eqn\iio{N_{\omega} \sim \o^{-a}e^{\omega/ T_H},~~~~ d{\cal E}_\o \sim \o^{p-1}
e^{-\o/T_H}e^{2 \pi X^0 T_H} d\o }
where the constant $a$ here is the number of non-compact
directions transverse to the brane.  
Evidently the integral \rlp\ diverges, except for sufficiently large numbers of 
non-compact transverse dimensions.\foot{This is analogous to 
'limiting' and 'non-limiting' Hagedorn behavior \rab.} The
total energy in open string production then receives an infinite contribution
from high-energy modes. 
Roughly speaking the brane tries to produce open strings 
at the Hagedorn temperature (although the final state is of course a pure
state). Since the particle production turns on slowly, there must be some
finite time $t_C$ at which the production rate diverges.\foot{Conceivably
this is related to the time at which the tachyon becomes massless.} $t_C$ is
independent of $g_s$.  

Of course when back reaction is included, this is forbidden by energy
conservation. The energy for the pair production must come out of the 
the tachyon background. No matter how small $g_s$ is 
the open strings become strongly coupled 
before $t=t_C$ and the linearized approximation breaks down. Eventually all
the energy goes into closed strings. It is hard to say exactly how this
occurs, but the divergence makes it plausible that the brane energy is
dumped into closed strings in a time of order $t_C$, despite the fact that 
closed strings formally decouple for $g_s \to 0$. In any case because 
of the divergence, the formal $g_s \to 0$ limit appears not to be smooth
in the minisuperspace approximation.

\centerline{\bf Acknowledgements} I am
grateful to M. Gutperle, A. Maloney, S. Minwalla, A. Recknagel, 
V. Schomerus, A. Sen and 
S. Shenker for
useful conversations. This work was supported in part by DOE grant
DE-FG02-91ER40654.

\listrefs

\end